\documentclass[aps,nofootinbib,preprintnumbers,longbibliography,amssymb,tightenlines,11pt,superscriptaddress]{revtex4} 
\pdfoutput=1
\usepackage[sort&compress]{natbib}
\usepackage{latexsym}
\usepackage{amsthm}
\usepackage{float}
\usepackage{amsmath,amssymb}
\usepackage{graphicx,subfigure,grffile}
\usepackage{feynmp}
\usepackage{indentfirst}
\usepackage{epsfig,subfigure,psfrag}
\usepackage{dcolumn}
\usepackage{bm}
\usepackage{slashed}
\usepackage{cancel}
\usepackage{color}
\usepackage{tabularx}
\usepackage[titletoc]{appendix}
\usepackage{hyperref}
\usepackage[usenames,dvipsnames]{xcolor}
\usepackage[normalem]{ulem}
\usepackage{color}
\usepackage{graphics}
\usepackage{float}
\usepackage{multirow}
\usepackage{amssymb}
\usepackage{array}
\usepackage{amsmath}

\begin{document}
\title{Revisiting Gluinos at LHC}

\preprint{LCTP-18-07}

\author{Malte Buschmann}
\email{buschman@umich.edu}
\affiliation{Leinweber Center for Theoretical Physics, Department of Physics, University of Michigan, Ann Arbor, MI 48109 USA}

\author{Eric Gonzalez}
\email{ericgz@umich.edu}
\affiliation{Leinweber Center for Theoretical Physics, Department of Physics, University of Michigan, Ann Arbor, MI 48109 USA}

\author{Gordon L. Kane}
\email{gkane@umich.edu}
\affiliation{Leinweber Center for Theoretical Physics, Department of Physics, University of Michigan, Ann Arbor, MI 48109 USA}

\begin{abstract}
	We examine the experimental signature of a UV complete Supersymmetry (SUSY) theory, the $G_2$-MSSM. This model predicts that only some superpartners will be produced in possibly detectable amounts at LHC: $p p \rightarrow \tilde{g}\tilde{g}$, $p p \rightarrow \tilde{\chi}^{\pm}_1 \tilde{\chi}^{\mp}_1$, and $p p \rightarrow \tilde{\chi}^{0}_2 \tilde{\chi}^{\pm}_1$. We exclude spectra with $m_{\tilde{g}} \approx1.5$ TeV. While spectra with $m_{\tilde{g}} \approx1.7$ TeV and $m_{\tilde{g}} \approx1.9$ TeV are currently allowed (contrary to what is often claimed), data in hand could exclude these spectra. This is not in tension with reported exclusion limits due to the difference in decay topologies between simplified models and a UV motivated ($G_2$-MSSM) model. 

\end{abstract}

\maketitle
\section{Introduction}
\indent

With Run 2 of the LHC underway, we are better equipped than ever to explore TeV scale physics. While simplified models have the advantage of exploring broad swathes of parameter space, they are unlikely to be compatible with a UV completion. The subject of this note is one such framework with a UV completion, the ``$G_2$-MSSM" \cite{Acharya:2006ia,Acharya:2007rc,Acharya:2008zi,Acharya:2008hi,Ellis:2014kla}. As a theory within the paradigm of Supersymmetry (SUSY), the $G_2$-MSSM does the job of addressing many problems that SUSY is associated with solving: the Hierarchy Problem, WIMP Dark Matter, Unification, and more. 

The $G_2$-MSSM is not arbitrary. It arises \cite{Acharya:2006ia,Acharya:2007rc,Acharya:2008zi} by compactifying M-theory on a 7 dimensional manifold of $G_2$ holonomy which automatically generates $\mathcal{N}=1$ supersymmetry. Then, a generic K\"ahler potential and a generic gauge kinetic function are used along with a superpotential consistent with the axion shift symmetry. All moduli are stabilized as supersymmetry is broken with F-terms nonzero at about $10^{14}$ GeV. The result is a gravitino mass of about 40 TeV. Ultimately, the theory has no free parameters, though some variation in the soft breaking masses and sparticle masses arises from computational uncertainties.

One of the key phenomenological features of this framework is that the SUSY particle (sparticle) spectrum has been approximately calculated with physically motivated high scale parameters~\cite{Ellis:2014kla}. Thus, with the discovery of just one sparticle, one can predict the rest of the mass spectrum. With this in mind we believe it to be important to explore the discovery potential of sparticles within this framework. In this note we will explore limits set by validated analyses on this model and extrapolate to high luminosities for future collider runs. In section 2 we discuss the aspects of the $G_2$-MSSM that we study at the LHC. In section 3 we discuss our method for analyzing the exclusion of our benchmark spectra. In section 4 we tabulate results and discuss concluding remarks.

\clearpage
\section{Collider Phenomenology of $G_2$-MSSM}
\indent

In this section we discuss the aspects of the $G_2$-MSSM relevant to the production of superpartners in high energy processes at the LHC. We use results from previous work \cite{Ellis:2014kla} and \texttt{SOFTSUSY}~\cite{Allanach:2001kg} to compute three benchmark SUSY spectra (Table 1). \texttt{SOFTSUSY} takes as input high scale parameters and runs down to collider scales using RGEs at the two loop level including threshold corrections. High scale parameters are drawn from a 3D parameter space of $M_{3/2}$ (gravitino mass), $\mu$ (Higgs bilinear term in the superpotential), and $C$ (a constrained parameterization of K\"ahler Potential corrections). These parameters are related to SUSY soft breaking parameters $M_a$ (high scale gaugino masses), $m_0$ (universal soft scalar mass) and $A_0$ (trilinear) in the following form~\cite{Acharya:2008hi}:

\begin{equation}
m_0^2 = M^2_{3/2}(1-C)
\end{equation}
\begin{equation}
A_0= 1.5 M_{3/2}(1-C)
\end{equation}
\begin{equation}
M_a = [-0.032\eta + \alpha_{{\text{GUT}}}\big(0.034\ (3C_a-C_a')+0.079C_a'(1-C)\big)]\times M_{3/2} 
\end{equation}
\begin{equation}
\tan{\beta} \approx \frac{M_{3/2}(1-C)}{2\mu}
\end{equation}
where $C_a=(0,2,3)$ and $C_a'=(33/5,7,6)$. We used these equations to compute soft breaking parameters, and \texttt{SOFTSUSY} to compute the spectra in Table 1. As the 3D parameter space is essentially constrained to 1D we can choose one of three parameters to characterize our spectra so long as we satisfy these constraints. We select values of $M_{3/2}$ to explore different spectra while $\mu$ and $C$ are selected to be consistent with EWSB and Higgs mass, computed with \texttt{SUSYHD} \cite{Vega:2015fna}. In this way we cover all relevant parameter space with these benchmark points.

\begin{table} [H]
	\centering
	\begin{tabular}{ c  c  c  c }
		\hline
		\multirow{3}{*}{\textbf{Spectrum}} 
		& $M_{3/2} = 30.8$ TeV&$M_{3/2} = 36.0$ TeV & $M_{3/2} = 41.8$ TeV \\
		&  $\mu=949$ GeV &  $\mu=1296$ GeV  &  $\mu=1747$ GeV \\
		& $C=0.5368$ & $C=0.5160$ & $C=0.4928$ \\
		\hline 
		
		Sparticle & \multicolumn{3}{c|}{Sparticle Masses [GeV]}  \\
		\hline 
		$\tilde{q}_{L,R}$ & 20800 & 25000 & 29500  \\
		
		$\tilde{l}_{L,R}$ & 20900 & 25000 & 29700  \\
		
		$\tilde{\tau}_{1,2}$, $\tilde{\nu}_{\tau L}$ & 20800 & 25000 & 29500 \\
		
		$\tilde{b}_1$ & 16700 & 20000 & 23800 \\
		
		$\tilde{b}_2$ & 20700 & 24700 & 29400 \\
		
		$\tilde{t}_1$ & 11500 & 13900 & 16500 \\
		
		$\tilde{t}_2$ & 16700 & 20000 & 23800 \\
		
		$\tilde{\chi}^0_4$ ($\tilde{H}^0_u$,$\tilde{H}^0_d$ mix)& 990 & 1340 & 1800 \\
		
		$\tilde{\chi}^0_3$ ($\tilde{H}^0_u$,$\tilde{H}^0_d$ mix)& 980 & 1340 & 1800 \\
		
		$\tilde{\chi}^0_2$ ($\tilde{W}^0$ like)& 590 & 690 & 780  \\
		
		$\tilde{\chi}^0_1$ ($\tilde{B}$ like)& 460 & 540 & 630 \\
		
		$\tilde{\chi}^\pm_2$ ($\tilde{H}^\pm_{u/d}$ like)& 990 & 1340 & 1800 \\
		
		$\tilde{\chi}^\pm_1$ ($\tilde{W}^\pm$ like)& 590 & 690 & 780 \\
		
		$\tilde{g}$ & 1490 & 1680 & 1890 \\
		\hline

	\end{tabular}
	\caption{Sparticle masses of tested spectra and relevant high scale parameters.}
\end{table}

In these spectra, scalars are larger than 10 TeV and we do not expect to produce them at the LHC. Another effect of heavy squarks is that the gluino cross section is  suppressed, and branching ratios are affected. Furthermore, we found that the heavier electroweak sparticles, $\tilde{\chi}^\pm_2$, $\tilde{\chi}^0_3$, and $\tilde{\chi}^0_4$ have LHC-13 cross sections which are far too small to be considered. This leaves us with four particles ($\tilde{\chi}^0_1$, $\tilde{\chi}^\pm_1$, $\tilde{\chi}^0_2$, and $\tilde{g}$ and, consequently, only three production channels ($p p \rightarrow \tilde{g}\tilde{g}$, $p p \rightarrow \tilde{\chi}^{\pm}_1 \tilde{\chi}^{\mp}_1$, and $p p \rightarrow \tilde{\chi}^{0}_2 \tilde{\chi}^{\pm}_1$) to search for SUSY at the LHC. Other light production channels  are not considered as production cross sections are small due to weak couplings (e.g. $\sigma(p p \rightarrow \tilde{\chi}^{0}_1 \tilde{\chi}^{\pm}_1)=0.06044 $ fb).

\begin{table}
	\centering
	\begin{tabular}{ c | c | c }

	Spectrum & \qquad\qquad Process \qquad\qquad\qquad & Cross Section [fb] \\ \hline
	&$p p \rightarrow \tilde{g}\tilde{g}$ & 14.0$\pm$0.9 \\ 
	$M_{3/2} = 30.8 $ TeV & $p p \rightarrow \tilde{\chi}^{\pm}_1 \tilde{\chi}^{\mp}_1$ & 8.2$\pm$0.6 \\
	&$p p \rightarrow \tilde{\chi}^{0}_2 \tilde{\chi}^{\pm}_1$ & 4.8$\pm$0.6 \\ \hline
	&$p p \rightarrow \tilde{g}\tilde{g}$ & 4.8$\pm$0.9 \\
	$M_{3/2} = 36.0$ TeV & $p p \rightarrow \tilde{\chi}^{\pm}_1 \tilde{\chi}^{\mp}_1$ & 4.2$\pm$0.6 \\
	&$p p \rightarrow \tilde{\chi}^{0}_2 \tilde{\chi}^{\pm}_1$ & 2.4$\pm$0.6 \\ \hline
	&$p p \rightarrow \tilde{g}\tilde{g}$ & 1.5$\pm$0.9 \\
	$M_{3/2} = 41.8 $ TeV &$p p \rightarrow \tilde{\chi}^{\pm}_1 \tilde{\chi}^{\mp}_1$ & 1.9$\pm$0.7 \\
	&$p p \rightarrow \tilde{\chi}^{0}_2 \tilde{\chi}^{\pm}_1$ & 1.2$\pm$0.6\\ 
	\end{tabular}
	\caption{Cross sections of interest for tested spectra from UV complete theory computed with \texttt{PROSPINO}.}
	
\end{table}

\begin{table}
	\centering
	\begin{tabular}{c | c }
		Decay & \qquad BR\qquad\qquad  \\
		\hline
		$\tilde{\chi}^{\pm}_1 \rightarrow W^{\pm} \tilde{\chi}^0_1$ & 100\% \\ \hline 
		$\tilde{\chi}^{0}_2 \rightarrow h \tilde{\chi}^0_1$ & 98\% \\
		$\tilde{\chi}^{0}_2 \rightarrow Z \tilde{\chi}^0_1$ & 2\% \\ \hline
		\boldmath $\tilde{g} \rightarrow Heavy \  Quarks$ & \boldmath $55\%$ \\\hline
		$\tilde{g} \rightarrow \tilde{\chi}^{\pm}_1 b\bar{t}, t\bar{b}$ & 23\% \\
		$\tilde{g} \rightarrow \tilde{\chi}^{0}_1 t\bar{t}$ & 20\% \\
		$\tilde{g} \rightarrow \tilde{\chi}^{0}_2 t\bar{t}$ & 4\% \\
		$\tilde{g} \rightarrow \tilde{\chi}^{0}_1 b\bar{b}$ & 1\% \\
		$\tilde{g} \rightarrow \tilde{\chi}^{0}_2 b\bar{b}$ & 7\% \\
		\hline 
		\boldmath $\tilde{g} \rightarrow Light \  Quarks$ & \boldmath $45\%$ \\ \hline
		$\tilde{g} \rightarrow \tilde{\chi}^+_1 q_{1,2}\bar{q}_{1,2}$ & 25\% \\
		$\tilde{g} \rightarrow \tilde{\chi}^0_2 q_{1,2}\bar{q}_{1,2}$ & 12\% \\
		$\tilde{g} \rightarrow \tilde{\chi}^0_1 q_{1,2}\bar{q}_{1,2}$ & 8\% \\ 
	\end{tabular}
	\caption{Sparticle Decay Branching Fractions \cite{Ellis:2014kla}.}
\end{table}

With these production channels, we should also examine decay branching fractions and cross sections (Tables 2 \& 3) to understand the type of analyses relevant to this model. One feature of this framework, which simplifies our work, is the fact that $\tilde{\chi}^\pm_1 \rightarrow \tilde{\chi}^0_1 W^\pm$ and $\tilde{\chi}^0_2 \rightarrow \tilde{\chi}^0_1 h$ are both essentially 100\% branching fractions (Table 3). Gluinos have a more complicated decay topology which forces us to consider several different search strategies. We note that although some analyses seem to exclude these parts of SUSY parameter space \cite{Aaboud:2017hrg,Aaboud:2017bac,Aaboud:2017vwy,Aaboud:2017dmy,Aaboud:2017hdf,Sirunyan:2017pjw,Sirunyan:2017bsh}, they do so with simplified models which do not account for more complex branching fraction topologies from UV theories like those in Table 3.

\clearpage
\section{Methods}
\indent

In this report we use \texttt{CheckMATE 2}~\cite{Dercks:2016npn} which compares a model against recast and validated 13 TeV analyses to set limits against existing data. We use rates predicted by the analyses recast in \texttt{CheckMATE 2} to predict future exclusion limits \& discovery potential. We use \texttt{MCLimit} \cite{Junk:1999kv} to calculate limits on rescaled rates.

The first step is to produce Monte Carlo (MC) events of our model to feed into \texttt{CheckMATE 2}. We begin by taking UV scale boundary conditions from previous work \cite{Ellis:2014kla} to compute spectra using \texttt{SOFTSUSY}. From this output we use \texttt{sDecay} \cite{Muhlleitner:2003vg} to compute mixing angles and branching fractions for a complete \texttt{SLHA} \cite{Skands:2003cj} model file. With this we are ready to generate events using the \texttt{MadGraph5} \cite{Alwall:2014hca} event generator interfaced with \texttt{PYTHIA 8.2} \cite{Sjostrand:2014zea} and \texttt{DELPHES 3} \cite{deFavereau:2013fsa} for appropriate hadronization and fast detector simulation. Events are normalized to NLO cross section computed by \texttt{PROSPINO2} \cite{Beenakker:1996ed} (Table 2).  With these events we can then run \texttt{CheckMATE2} to see how this model compares against current 13 TeV ATLAS analyses (CMS analyses have not yet been validated in \texttt{CheckMATE2}). From there, we estimate the luminosity required to discover or exclude this model by scaling rates linearly to find limits with \texttt{MCLimit}.

\section{Conclusion}
\indent

The results of this process are summarized in Table 4, where we have selected the two best signal regions for exclusion from the \texttt{CheckMATE2} report. To clarify, $\mathcal{L}$ represents the luminosity at which the respective analysis was performed, while $\mathcal{L}^{95}$ is the luminosity required to exclude the spectrum in a given model with rescaled rates from this analysis. $N_{obs}$, $N_b$, and $N_s$ are the observed (data), background (MC) and signal (MC) event numbers after performing the respective analysis on a given data set.

\begin{table}[H]
\centering
\resizebox{\columnwidth}{!}{%
\begin{tabular}{ c | c | c | c | c | c | c | c  }

		 Analysis & $\mathcal{L}$ [fb$^{-1}$] & Spectrum & Signal Region & $N_{obs}$ & $N_b$ & $N_s$  & $\mathcal{L}_{95}$ [fb$^{-1}$]\\
		\hline 
		\multirow{3}{*}{ATLAS 1605.09318}
		& 					 	 & $m_{\tilde{g}} = 1.5$ TeV  & SR Gtt-1L-B&  0 &  1.2 & 5.1  &   $<3.3$ \\ 
		& 3.3    & $m_{\tilde{g}} = 1.7$ TeV& SR Gtt-1L-B & 0 & 1.2 & 2.0 &  11.0 \\ 
		&						 & $m_{\tilde{g}} = 1.9$ TeV  & SR Gtt-0L-A  & 1  & 2.1  & 0.8   & 82.5  \\ \hline

		\multirow{3}{*}{ATLAS 1709.04183}
		& 						 & $m_{\tilde{g}} = 1.5$ TeV & SRD-low & 27 & 25.1 & 24.6 & $<36.1$ \\ 
		& 36.1 & $m_{\tilde{g}} = 1.7$ TeV  & SRA-TT    & 11   & 8.6   & 6.4    &  66.8 \\ 
		& 						 & $m_{\tilde{g}} = 1.9$ TeV & SRA-TT  & 11  & 8.6  & 2.2    & $>300$ \\

	\end{tabular}
	}
	
	\caption{Best signal regions from \texttt{CheckMATE2}, and luminosity projections for 95\% CL. exclusion.}
	\end{table}
\indent

The results indicate that some $G_2$-MSSM spectra within the region of interest derived from previous work \cite{Ellis:2014kla} are already excluded. Namely, we tested three spectra which we characterize 
by the gluino mass $m_{\tilde{g}}$ with values 1.5, 1.7, and 1.9 TeV. We found that the $m_{\tilde{g}}\approx1.5$ TeV spectrum is indeed excluded by data while the $m_{\tilde{g}} \approx 1.7$ TeV and $m_{\tilde{g}}  \approx1.9$ TeV benchmark spectra are not currently excluded by existing analyses, but are expected excluded or discovered to be as searches are updated.

The two most effective searches for excluding these models, ATLAS 1605.09318 and ATLAS 1709.04183, target gluino decays to third family quarks with large amounts of missing energy. In fact, the best signal regions were ones which looked for top quarks and their decay products (Gtt and SRA/D type signal regions). Although electroweak sparticle production channels have comparable cross sections to gluino production, the larger mass splitting between $\tilde{g}$ and $\tilde{\chi}^0_1$ makes gluino searches much more favorable as the mass difference allows for hard $p_T$ jets. To be precise, according to our results, an analysis which searches for gluino decays to top quarks with 0 leptons and moderate mass splittings (Gtt-0L-B) could exclude the $m_{\tilde{g}} = 1.7$ TeV model with as little as 11.0 fb$^{-1}$ LHC-13 data (which has been collected). The search for a $m_{\tilde{g}} = 1.9$ model sharpens by including a lepton in the search (Gtt-1L-A) and could be found with about 82.5 fb$^{-1}$ LHC-13 data (which is expected to be collected soon).

Finally, we emphasize that our results are not inconsistent with current ATLAS and CMS results as simplified models assume 100\% branching rates to heavy family quarks. UV complete models, however, have a more complex decay topology. In the case of the $G_2$-MSSM gluinos will decay to third family quarks roughly half the time, and first and second family quarks the other half  (Table 3) which accounts for the reduction in signal efficiency. This simple physical property tells us that exclusion limits should be scaled down when considering more than simplified models, and that exclusion limits on sparticles are in fact more nuanced than we are commonly led to believe. Although we only tested a specific set of models within a UV complete framework, a lesson learned in this exercise is that quoted SUSY limits are, at face value, spurious. One should carefully consider the assumptions made in setting a limit and be sure not to apply them inappropriately. We encourage experiments to additionally quote limits on a set of realistic models so that readers could have a better understanding of how simplified models should be interpreted.

\section*{Acknowledgments}
Thanks to Sebastian A.R. Ellis for helpful discussion. Work of G.L.K. and E.G. is supported by Department of Energy grant DE-SC0007859.

\bibliographystyle{JHEP}
\bibliography{bibtexrefs}

\end{document}